\documentclass[amsmath,amsfonts,amssymb,prl,twocolumn,superscriptaddress,showpacs]{revtex4}

\usepackage{graphicx,psfrag}
\usepackage{dcolumn}
\usepackage{bm}
\usepackage{mathbbol}

\newcommand{\br}{{\bf r}}

\begin{document}

\title{\mbox{Order-Parameter Correlation Functions in Quantum Critical Phenomena}}

\author{Min-Chul Cha} 
\affiliation{\mbox{Department of Applied Physics, Hanyang University,
Ansan, Kyunggi-do 426-791, Korea}}
\author{Gerardo Ortiz}
\affiliation{Department of Physics, Indiana University, Bloomington,
IN 47405, USA}

\date{\today}

\begin{abstract}
We investigate the functional form of the order-parameter (two-point) 
correlation function in quantum critical phenomena. Contrary to the 
common lore, when there is no particle-hole symmetry we find that the
equal-time correlation function at criticality does not display a
diverging correlation length. We illustrate our conclusions by Monte
Carlo calculations of the quantum rotor model in $d=2$ space
dimensions.
\end{abstract}

\maketitle


To probe the dynamical response of an interacting quantum many-body
system one typically investigates its response to an externally applied
time-dependent force that perturbs the system slightly away from 
equilibrium. The two-point time-dependent correlation function (TPCF)
is the quantity which encodes the physical response of the system to
that perturbation in linear response theory \cite{martin}. In this way,
the TPCF contains information about the nature of the excitations of
the unperturbed system. If the system is close to a continuous
(thermodynamic or quantum) phase transition, under certain assumptions
(e.g. long-distances), the order-parameter TPCF may acquire a universal
form independent of the microscopic details of the interactions. This
is the beauty and power of the general theory of critical phenomena.  
Establishing this universal functional form, and determining the
exponents involved, is a common theme in classical and quantum critical
phenomena \cite{goldenfeld}. 

Quantum Critical Phenomena has been introduced as an  straightforward
extension of Classical Critical Phenomena \cite{hertz}. The motivation
is clear and can be traced back to the formal functional-integral
mapping between a quantum problem in $d$ space dimensions and a
classical statistical mechanics problem in $d+1$, where the extra
dimension corresponds to the (imaginary)  time axis $\tau$ (see, for
example, \cite{parisi}).  Space and time coordinates do not necessarily
satisfy Lorentz symmetry. A measure of the asymmetry is quantified by
the dynamic critical exponent $z$ which is equal to unity when Lorentz
invariance is present. Common wisdom dictates that {\it at the critical
point} $g_c$ continuous phase transitions display a correlation length
$\xi$ diverging as $\xi \sim  |g-g_c|^{-\nu}$ with critical exponent
$\nu>0$,  and control parameter $g$ (temperature in the case of
thermodynamic  phase transitions). Similarly, the correlation time
diverges as $\tau_c \sim \xi^z$. This fact reflects the scale
invariance of the system at criticality. It is then emphasized that
consequently the {\it equal-time} (order-parameter) TPCF $G(\br,
\tau-\tau'=0)$ should decay algebraically with the distance $r=|\br|$,
$G(\br,0) \sim 1/r^{d+z-2+\eta}$, thus defining a characteristic
exponent  $\eta$ (anomalous dimension). Different relations or {\it
laws}  are satisfied by the various exponents as a result of {\it
scaling hypotheses}  (e.g. in the free energy) \cite{cardy}. 

The purpose of this paper is to note that the common lore depicted
above is not always mathematically justified in quantum critical
phenomena and, most importantly, inadequate in certain cases of
physical relevance. The root of the discrepancy lies in the lack of
particle-hole symmetry in the case of non-vanishing chemical potential
$\mu$, which leads to a non-divergent length at criticality and
short-range equal-time correlations as determined from $G(\br,0)$. This
in turn implies that when $\mu \neq 0$ there is no power law scaling
ansatz in $G(\br,0)$ at the critical point. On the other hand, the {\it
zero-frequency} order-parameter correlation function, $G(\br,
\omega=0)$, does display long-range power law behavior at criticality
with a divergent correlation length $\xi$ for arbitrary $\mu$. This
illustrates the fundamental physical distinction between equal-time and
zero-frequency correlations.  We will present both a theoretical proof
of these statements within the Ginzburg-Landau-Wilson (GLW) paradigm of
quantum phase transitions, with an analytic calculation for the quantum
spherical model, and a  numerical confirmation in the quantum rotor
model. Our conclusions rigorously  apply as long as the dynamics of the
order parameter of a given microscopic quantum model is  faithfully 
described by this kind of GLW effective field theory. This holds
regardless of  the nature of the  original degrees of freedom, i.e 
whether bosonic or fermionic. Moreover, similar  derivations can be
performed with multicomponent order parameters. 

Our starting quantum model is of the familiar phenomenological GLW
form 
\begin{widetext}
\begin{eqnarray}
S[\psi,\psi^*]&=&\frac{1}{2}\int_0^{1/T} d\tau \int_V d^d{\bf r}
\big\{-\psi^*({\bf r,\tau})
[(\partial_\tau-\mu)^2+\nabla_\br^2]\psi({\bf r,\tau}) +r_0|\psi({\bf
r},\tau)|^2+\frac{u_0}{2}|\psi({\bf r},\tau)|^4\big\}\ ,
\label{eq:GLW}
\end{eqnarray}
\end{widetext}
where $\psi(\br,\tau)$ is the space and (Euclidean) time-dependent
complex order parameter field, $r_0$ and $u_0$ are physical parameters,
$V=L^d$ is the volume, and $T$ is the temperature. The partition
function is given by ${\cal Z}=\int {\cal D}[\psi]{\cal D}[\psi^*]
\exp(-S[\psi,\psi^*])$. This (Euclidean) action $S$ is symmetric under
the transformation $\psi(\br,\tau)\rightarrow -\psi(\br,\tau)$, but it
is not particle-hole symmetric because $\mu$ is, in general, not zero. 
This GLW functional represents an analytic expansion (in the vicinity
of the critical point) in  $\psi(\br,\tau)$ with interaction terms
respecting the group of symmetries ${\cal G}$ of the  disordered
phase. 

Under the assumption of (bosonic) periodic boundary conditions along
the time axis, one can Fourier transform the fields as
\begin{eqnarray}
\psi({\bf r},\tau)=\sqrt{\frac{T}{V}}\sum_{{\bf k},\omega} e^{i({\bf
k}\cdot{\bf r}-\omega\tau)}\tilde\psi({\bf k},\omega) ,
\end{eqnarray}
with Matsubara frequencies $\omega=2 \pi T n \ (n=0,\pm 1,\pm
2,\cdots)$, and discrete momenta ${k}_\alpha=2 \pi p_\alpha/L \
(p_\alpha=0,\pm 1,\pm 2,\cdots)$ with $\alpha=1,\cdots,d$.  Then, 
\begin{eqnarray}
S&=&\frac{1}{2}\sum_{{\bf k},\omega}
[(\omega-i\mu)^2+k^2+r_0]|\tilde\psi({\bf k},\omega)|^2 \nonumber
\\
&+&\frac{u_0}{4}\int_0^{1/T} d\tau \int_V d^d{\bf r} |\psi({\bf
r},\tau)|^4 .
\end{eqnarray}
Introducing an auxiliary scalar field $\phi({\bf r},\tau)$, we 
decouple the last term by the Hubbard-Stratonovich identity 
\begin{eqnarray}
&& \hspace*{-0.7cm} e^{-\frac{u_0}{4}\int d\tau \int d^d{\bf
r} |\psi({\bf r},\tau)|^4 }={\rm const.} \nonumber \\
&& \!\!\!\!\!\! \times \int {\cal D}[\phi]
e^{-\frac{1}{2}\int d\tau \int d^d{\bf r} [i \phi({\bf
r},\tau) |\psi({\bf r},\tau)|^2+\frac{1}{2u_0} \phi^2({\bf r},\tau)]} ,
\end{eqnarray}
with the result for the modified action $\bar{S}[\psi,\psi^*,\phi]$
\begin{eqnarray}
\!\!\!\!\!\!\bar{S}&=&\frac{1}{2}\sum_{{\bf k},\omega}
[(\omega-i\mu)^2+k^2+r_0]|  \tilde\psi({\bf k},\omega)|^2 \nonumber \\
&+&\!\! \frac{1}{2}\! \int_0^{1/T} \!\!\!\!\!\!d\tau \int_V \!\! d^d{\bf r} 
[\frac{1}{2u_0}\phi^2({\bf r},\tau) +i\phi({\bf r},\tau)|\psi({\bf
r},\tau)|^2].
\end{eqnarray}
In order to investigate the order-parameter TPCF explicitly, we will
adopt a saddle-point approximation to evaluate the integral over $\phi$.
The resulting (quantum spherical model) action is
\begin{eqnarray}
\tilde{S}&=&\frac{1}{2}\sum_{{\bf k},\omega}
[(\omega-i\mu)^2+k^2+\bar{r}_0]|  \tilde\psi({\bf k},\omega)|^2 ,
\nonumber
\end{eqnarray}
where $\bar{r}_0=r_0+i \bar{\phi}$.  For the sake of clarity, in the
following we will concentrate in the $T=0$ and infinite volume case.

The resulting self-consistent equation is
\begin{eqnarray}
i\bar\phi
=u_0\int_{-\infty}^{\infty}\frac{d\omega}{2\pi}\int  \frac{d^d{\bf
k}}{(2\pi)^d} \frac{1}{[(\omega-i\mu)^2+k^2+\bar{r}_0]} . 
\end{eqnarray}
Equivalently, we can define a correlation length $\xi$
\begin{eqnarray}
\xi^{-2}= r_0-\mu^2+u_0\int\frac{d^d{\bf k}}{(2\pi)^d}
\frac{1}{2\sqrt{k^2+\mu^2+\xi^{-2}}} ,
\label{eq:SCE}
\end{eqnarray}
where $\xi^{-2}\equiv r_0+i\bar\phi-\mu^2$. Clearly, this integral
displays an ultraviolet divergence that needs to be regularized. To
avoid this divergence, here we adopt a lattice regularization by
replacing $k^2 \to \sum_{\alpha}2(1-\cos k_\alpha)$ and
$\int\frac{d^d{\bf k}}{(2\pi)^d} \to (1/V)\sum_{\bf k}$, which
correctly takes into account the long-wavelength contributions while
providing an intrinsic cutoff due to the lattice constant.

\begin{figure}[b]
\includegraphics[width=0.94\columnwidth]{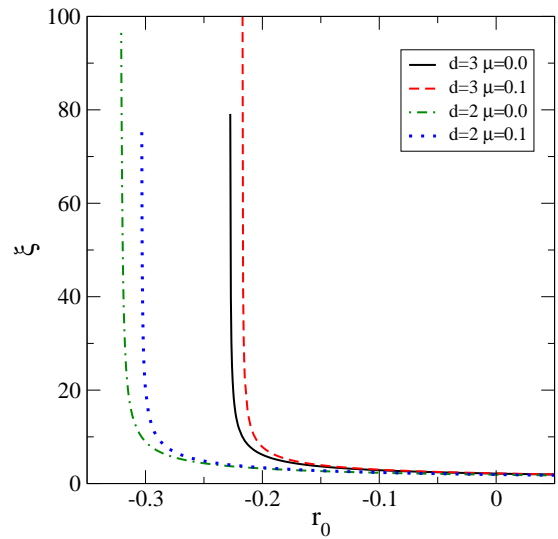}
\caption{(color online)
The correlation length, $\xi$, determined by the self-consistent
Eq. (\ref{eq:SCE}), diverges as one approaches the transition 
point by tuning $r_0$. Here we consider the three- and two-dimensional
cases. ($u_0=1$)
}
\label{fig:xi}
\end{figure}

Figure~\ref{fig:xi} shows the behavior of $\xi$ in two and three
dimensions, where, without loss of generality, we set $u_0=1$ and use
$r_0$ as the control parameter to signal the quantum phase transitions
for $\mu=0$ and $\mu\ne 0$. We see that $\xi$ diverges as one
approaches the quantum critical point. Below we show that $\xi$ is
actually the correlation length, which gives the diverging (or
singular) behavior at criticality, but {\it it is not} the length
characterizing the TPCF.

With this correlation length $\xi$, the order-parameter (connected)
TPCF is given by
\begin{eqnarray}
G({\bf r},\tau)&=&\langle\psi({\bf r},\tau)\psi^*(0,0)\rangle
-\langle\psi({\bf r},\tau)\rangle\langle\psi^*(0,0)\rangle\cr
&=&
\int \frac{d^d{\bf k}d\omega}{(2\pi)^{d+1}} \frac{e^{i({\bf k}\cdot{\bf
r}-\omega\tau)}}{(\omega-i\mu)^2+k^2+\mu^2+\xi^{-2}}\cr
&=&
\frac{e^{\mu\tau}\lambda^{(1-d)}}{(2\pi)^{(d+1)/2}}
\rho^{-(d-1)/2}K_{(d-1)/2}(\rho)
\end{eqnarray}
for $d \ge 1$, where $\lambda^{-2}\equiv\mu^2+\xi^{-2}$,
$\rho=(r^2+\tau^2)^{1/2}/\lambda$, and $K_n(\rho)$ is the modified
spherical Bessel function of the third kind whose asymptotic form is
\begin{eqnarray}
K_n(\rho)\sim
\left\{\begin{array}{ll}
\big(\frac{\pi}{2\rho}\big)^{1/2}e^{-\rho} & \rho \gg 1;\\
\frac{\Gamma(n)}{2}\big(\frac{\rho}{2}\big)^{-n} & \rho \ll 1,
\end{array}\right.
\end{eqnarray}
where $\Gamma(n)$ is the Gamma function. Therefore, for $\mu \ne 0$,
the long-range behavior of the TPCF near the critical point
($\xi^{-2}\to 0$) reduces to
\begin{eqnarray}
G({\bf r},\tau) \sim
\frac{e^{-\mu(\sqrt{r^2+\tau^2}-\tau)}}{(r^2+\tau^2)^{d/4}}\ ,
\label{eq:cor}
\end{eqnarray}
while, for $\mu=0$, $G(\br,\tau)\sim 1/(r^2+\tau^2)^{(d-1)/2}$. We note
that, for $\mu \ne 0$, the equal-time spatial correlation shows a
short-range behavior characterized by a length $\lambda \sim 1/\mu$,
not by the diverging $\xi$, while the temporal correlation does display
long-range behavior.

The long-range correlation associated with the divergence of $\xi$,
however, leads to the susceptibility thermodynamic sum rule
\cite{goldenfeld} so that $\tilde G({\bf
k},\omega)=1/[(\omega-i\mu)^2+k^2+\mu^2+\xi^{-2}]\to \infty$ as
$\xi\to\infty$ in the long-wavelength ($k\to 0$) and low-frequency
($\omega\to 0$) limit. Therefore, the range of the zero-frequency
correlation function is characterized by $\xi$ in the form
\begin{eqnarray}
G({\bf r},\omega=0) \sim
\left\{\begin{array}{ll}
\exp(-r/\xi)/r^{(d-1)/2} & r/\xi \gg 1;\\
1/r^{d-2} & r/\xi \ll 1,
\end{array}\right.
\end{eqnarray}
which, at criticality, is often compactly represented as $G({\bf
r},\omega=0) \sim \exp(-r/\xi)/r^{d-2}$. Note that this form is the
same as that of the $d$-dimensional classical criticality. We may view
that the long-range spatial correlation at zero-frequency is brought by
the long-range temporal correlation, even though the equal-time
correlations are short ranged.

We see that in the model a non-vanishing $\mu$ brings in an imaginary
term linearly coupled to $\omega$ in the action. Without this term, the
quantum fluctuations along the temporal and spatial directions are 
isotropic, yielding a dynamical critical exponent $z=1$, and the
mapping of the $d$-dimensional quantum phase transition to the
$(d+z)$-dimensional classical phase transition is well justified. For
$\mu \ne 0$, the term breaking the particle-hole symmetry causes an
asymmetry in the temporal direction leading to $z=2$ (i.e. there is no
Lorentz invariance). This asymmetry property should not change at
higher orders in $u_0$. The formal solution, including the effect of
$u_0$, is
\begin{eqnarray}
\tilde G({\bf k},\omega)= \frac{1}{(\omega-i\mu)^2+k^2+r_0-
\Sigma(r_0,u_0,\mu; {\bf k},\omega)},
\end{eqnarray}
where the self-energy $\Sigma(r_0,u_0,\mu; {\bf k},\omega)\to 0$ as
$u_0\to 0$ and $\Sigma(r_{0c},u_0,\mu; 0,0) = r_{0c}-\mu^2$ at the
critical point $r_0=r_{0c}$. Therefore, we expect that the self-energy
term does not alter the nature of the broken particle-hole symmetry
caused by a non-zero $\mu$, although it may introduce an anomalous
exponent $\eta \neq 0$.

Our analytic results are now confirmed by a Monte Carlo calculation of
the quantum rotor model
\begin{eqnarray}
H=\frac{U}{2}\sum_{\bf r}
(\frac{1}{i}\frac{\partial}{\partial\theta_{\bf r}}-\bar n)^2
-2J\sum_{\langle {\bf r}, {\bf r^\prime}\rangle}\cos(\theta_{\bf
r}-\theta_{\bf r^\prime}),
\end{eqnarray}
where $U$ represents the rotational energy scale of a rotor at site
$\bf r$, $J$ denotes the coupling strength between nearest-neighbor
sites  $\langle {\bf r}, {\bf r^\prime}\rangle$, and a non-zero $\bar
n$ $(\bar n\equiv \mu/U)$ breaks the particle-hole symmetry.  In the
following, we will show that the correlations of this model near
criticality are correctly captured by the effective (GLW) model of Eq.
(\ref{eq:GLW}).

We performed Monte Carlo calculations of the total TPCF $G_{\rm
t}({\bf r},\tau)=\langle\psi({\bf r},\tau)\psi^*(0,0)\rangle =\langle
e^{i[\theta ({\bf r},\tau)-\theta (0, 0)]}\rangle$  in a square lattice
of size $L \times L \times L_\tau$, with periodic boundary conditions,
where $L$ is the size in a spatial direction and $L_\tau$ in the
temporal direction. For Monte Carlo calculations, we represent the
quantum rotor model in the form of a classical $(d+1)$-dimensional
action \cite{Wallin94}
\begin{eqnarray}
S[{\bf J}] =  \frac{1}{2K}\!\!\sum_{({\bf r},\tau)}^{\nabla\cdot{\bf
J}=0} \! \big\{J^2_{x({\bf r},\tau)} + J^2_{y({\bf r},\tau)} + (J_{\tau
({\bf r},\tau)}-\bar n)^2\big\},
\label{eq:cl_hamil}
\end{eqnarray}
where $K \sim \sqrt{2J/U}$ is the tuning parameter controlling the
quantum fluctuations, $J_\alpha$'s are integers, and ${\nabla\cdot{\bf J}=0}$
represents the current conservation condition at each site.  We use a
recently developed worm algorithm \cite{Alet03} to update the current
configurations in Eq.~(\ref{eq:cl_hamil}) and measure the correlation
functions.

\begin{figure}[t]
\includegraphics[width=0.94\columnwidth]{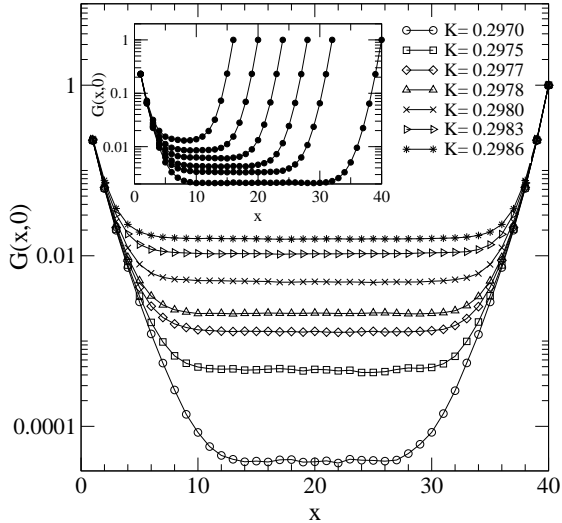}
\caption{
The TPCF $G(x,0)$ in a lattice of size $40\times 40\times 400$ for
different $K$'s near the critical point $K_c= 0.2978$ for $\mu/U=0.20$. The
characteristic length $\ell$, governing the range of the correlation
function, has a finite value and varies little for different $K$'s and
sizes. Inset: The same functions at $K=K_c$ in lattices of size $16
\times 16\times 64, 20\times 20\times 100, 24\times 24\times 144,
28\times 28\times 196, 32\times 32\times 256$, and $40\times 40\times
400$.
}
\label{fig:xcor}
\end{figure}

Figure~\ref{fig:xcor} shows the TPCF  as a function of $x$, a distance
along a spatial axis, in a lattice of size $40\times 40\times 400$ in
the vicinity of the critical point $K_c=0.2978$, which is determined by
finite-size scaling of the superfluid stiffness, for $\mu/U=0.20$. Inspired by
Eq.~(\ref{eq:cor}), we use the fitting function
\begin{eqnarray}
G_{\rm t}(x,\tau=0)=C+A \Big(\frac{e^{-x/\ell}}{x}+
\frac{e^{-(L-x)/\ell}}{(L-x)}\Big),
\end{eqnarray}
for $1 \ll x \ll L $ and obtain $\ell \approx 1.4 - 1.9$ for different
$K$'s and sizes, while $C=|\langle \psi\rangle|^2$ depends sensitively
on $K$. This result strongly supports the fact that the equal-time TPCF
has a short-range behavior, characterized by a finite  rather than a
diverging correlation length.

\begin{figure}[t]
\includegraphics[width=0.94\columnwidth]{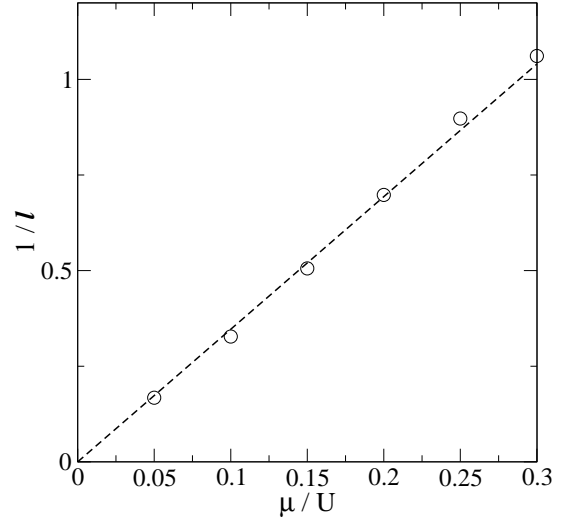}
\caption{
The inverse length, $1/\ell$, at the critical point ($\xi^{-1}=0$)
displays the behavior $1/\ell \propto \mu/U$. Calculations are
performed in a lattice of size $40\times 40\times 400$ at $K_c=0.3286,
0.3205, 0.3102, 0.2978, 0.2830$, and $0.2655$ for $\mu/U=0.05, 0.10,
0.15, 0.20, 0.25$, and $0.30$, respectively. The dotted line is a guide
to the eye.
}
\label{fig:inv_ell}
\end{figure}

\begin{figure}[t]
\includegraphics[width=0.94\columnwidth]{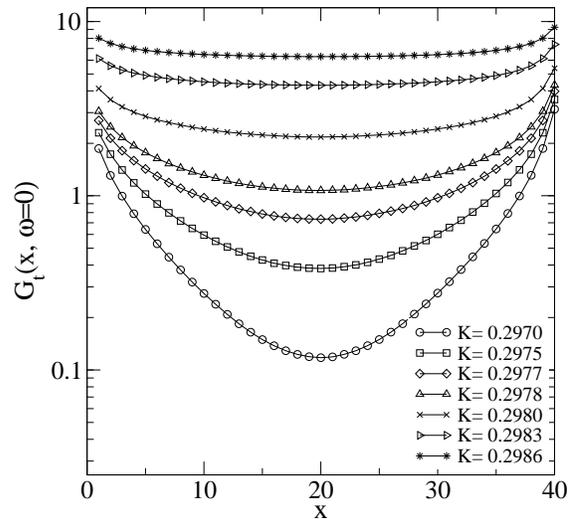}
\caption{
Zero-frequency correlation function near the critical point.
Long-range correlations are clearly observed.
}
\label{fig:xcor_w0}
\end{figure}

To understand the nature of $\ell$, we measure it as a function of
$\mu$ at the critical point in Fig.~\ref{fig:inv_ell}. Our results
clearly show that $1/\ell$ is proportional to $\mu$, and $\ell$ remains
finite except at integer fillings ($\mu=0$) where the
particle-hole symmetry is restored.

The long-range nature of one-particle correlations can be observed in
$G({\bf r}, \omega=0)$. In Fig. \ref{fig:xcor_w0}, the zero-frequency
total correlation function, defined by $G_{\rm t}(x,\omega=0) \equiv
\sum_{\tau=1}^{L_\tau}G_{\rm t}(x,\tau)$, displays long-range behavior
near the critical point. We can easily see that, by subtracting out the
constant part to obtain $G({\bf r}, \omega=0)$, the correlation range
is still beyond the size of the system.

In summary, we have shown that, when particle-hole symmetry is broken,
the physical quantity that displays a diverging length at criticality
is the {\it zero-frequency} and not the equal-time correlation
function.  The latter is short-ranged and within our paradigm of
quantum phase transitions it is not Lorentz invariant.  We have not
only proved this statement analytically but also illustrated its
correctness by studying the planar quantum rotor model.  This model
shares the same critical properties as the Bose-Hubbard model
\cite{Fisher89} assuming that only phase fluctuations are responsible
for its critical behavior. This observation is quite relevant given the
current interest in the experimental simulation of quantum phase
transitions in optical lattices of ultracold atomic gases
\cite{cold_atom}. 

M.-C.C. thanks the hospitality of the Department of Physics at Indiana
University where parts of this work were carried out, and the support by
the Korea Research Fund Grant No. KRF-2005-041-I01625.

\end{document}